\PassOptionsToPackage{table,xcdraw}{xcolor}
\documentclass[sigconf,authorversion]{acmart}

\usepackage{graphicx}
\usepackage{mathtools}
\usepackage{dsfont}
\usepackage{multirow}
\usepackage{caption}
\usepackage{subcaption}
\usepackage{todonotes}

\usepackage[multiple]{footmisc}

\AtBeginDocument{%
  \providecommand\BibTeX{{%
    \normalfont B\kern-0.5em{\scshape i\kern-0.25em b}\kern-0.8em\TeX}}}

\copyrightyear{2022}
\acmYear{2022}
\setcopyright{acmlicensed}\acmConference[SSDBM 2022]{34th International Conference on Scientific and Statistical Database Management}{July 6--8, 2022}{Copenhagen, Denmark}
\acmBooktitle{34th International Conference on Scientific and Statistical Database Management (SSDBM 2022), July 6--8, 2022, Copenhagen, Denmark}
\acmPrice{15.00}
\acmDOI{10.1145/3538712.3538739}
\acmISBN{978-1-4503-9667-7/22/07}

\begin{document}

\title{Lotaru: Locally Estimating Runtimes of Scientific Workflow Tasks in Heterogeneous Clusters}

\author{Jonathan Bader}
\affiliation{%
  \institution{Technische Universität Berlin}
  \city{Berlin}
  \country{Germany}
}
\email{jonathan.bader@tu-berlin.de}

\author{Fabian Lehmann}
\orcid{0000-0003-0520-0792}
\affiliation{%
   \institution{Humboldt-Universität zu Berlin}
  \city{Berlin}
  \country{Germany}}
\email{fabian.lehmann@informatik.hu-berlin.de}

\author{Lauritz Thamsen}
\affiliation{%
 \institution{University of Glasgow}
 \city{Glasgow}
 \country{United Kingdom}
 }
 \email{lauritz.thamsen@glasgow.ac.uk}

\author{Jonathan Will}
\affiliation{%
  \institution{Technische Universität Berlin}
  \city{Berlin}
  \country{Germany}
}
\email{will@tu-berlin.de}

\author{Ulf Leser}
\orcid{0000-0003-2166-9582}
\affiliation{%
  \institution{Humboldt-Universität zu Berlin}
  \city{Berlin}
  \country{Germany}}
  \email{leser@informatik.hu-berlin.de}

\author{Odej Kao}
\affiliation{%
  \institution{Technische Universität Berlin}
  \city{Berlin}
  \country{Germany}
}
\email{odej.kao@tu-berlin.de}

\renewcommand{\shortauthors}{Bader et al.}

\begin{abstract}
Many scientific workflow scheduling algorithms need to be informed about task runtimes a-priori to conduct efficient scheduling.
In heterogeneous cluster infrastructures, this problem becomes aggravated because these runtimes are required for each task-node pair.
Using historical data is often not feasible as logs are typically not retained indefinitely and workloads as well as infrastructure changes.
In contrast, online methods, which predict task runtimes on specific nodes while the workflow is running, have to cope with the lack of example runs, especially during the start-up.

In this paper, we present \emph{Lotaru}, a novel online method for locally estimating task runtimes in scientific workflows on heterogeneous clusters.
Lotaru first profiles all nodes of a cluster with a set of short-running and uniform microbenchmarks.
Next, it runs the workflow to be scheduled on the user's local machine with drastically reduced data to determine important task characteristics.
Based on these measurements, Lotaru learns a Bayesian linear regression model to predict a task's runtime given the input size and finally adjusts the predicted runtime specifically for each task-node pair in the cluster based on the micro-benchmark results.
Due to its Bayesian approach, Lotaru can also compute robust uncertainty estimates and provides them as an input for advanced scheduling methods.

Our evaluation with five real-world scientific workflows and different datasets shows that Lotaru significantly outperforms the baselines in terms of prediction errors for homogeneous and heterogeneous clusters.

\end{abstract}



\maketitle
\section{Introduction}\label{sec:INTRO}
Scientists from many domains, such as bioinformatics, remote sensing, and physics, use scientific workflow management systems to define, compose, and reproducibly execute their data analysis pipelines over large datasets ~\cite{deelman2019evolution, witt2019feedback, nextflow}.
These workflows are commonly organized as a directed acyclic graph (DAG), consisting of a set of abstract tasks T and a set of directed edges E.
While the abstract tasks serve as templates for their physical instances on real datasets, edges describe the flow of data between tasks and thus constrain the order of execution and degree of parallelism of task executions.

\begin{figure}[]
    \includegraphics[width=\columnwidth]{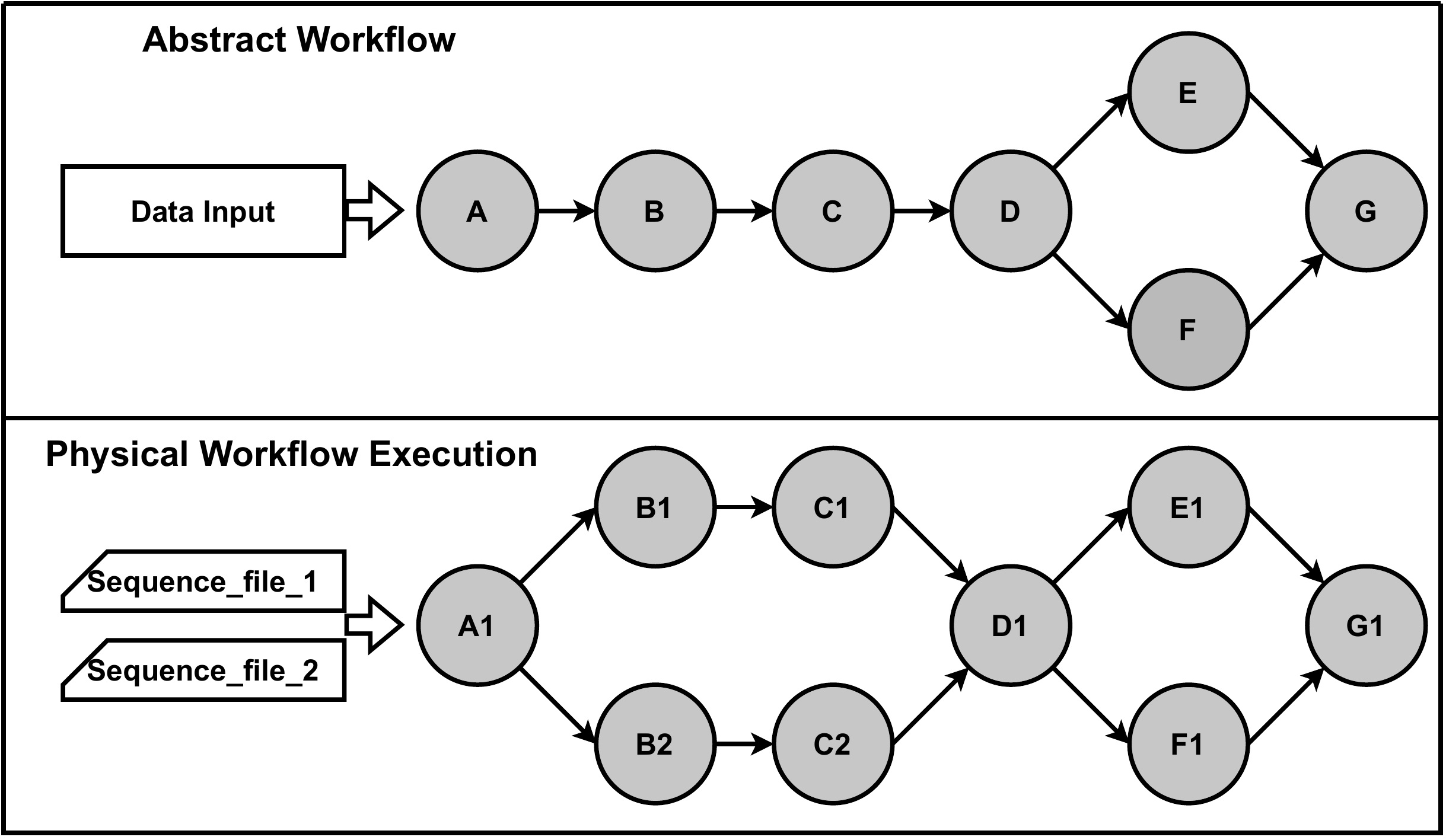}
    \caption{Abstract and physical execution model of a scientific workflow.}
	\label{fig:execution_model}
\end{figure}

Fig. \ref{fig:execution_model} shows an exemplary abstract workflow and a concrete physical execution.
The physical representation with two input data samples shows that the physical tasks B1, B2 and C1, C2, respectively, may be executed in parallel as they have no interdependencies.
Tasks D, E, F, and G result in only one physical task.

When such workflows are executed over large amounts of data, their runtimes can easily exceed days or even weeks~\cite{cybershake,da2020characterizing, bader2021tarema, lehmannFORCENextflowScalable2021}.
To effectively use the available cluster resources, many workflow management systems have a scheduling component~\cite{pegasus, saasfee, oinn2004taverna, koster2012snakemake} that determines which tasks are executed when and on which of the available nodes to achieve some optimization goal, such as wallclock-time.
To this end, most practical methods today require accurate estimates of the runtime of any task on any node~\cite{heft, pheft}.
As these are difficult to obtain, schedulers often resort to user configurations, which, however, are known to be highly error-prone~\cite{witt2019learning, witt2019predictive}.
In practice, the problem is complicated further because clusters available to scientists for their data analysis often consist of heterogeneous hardware~\cite{cpuperformance, bader2021tarema}.
Reasons are, for instance, partially upgraded nodes, hardware replacements over time, or clusters that are intended to serve multiple purposes.
In such settings, nodes' basic resource properties differ, like the size and speed of main memory, number and frequency of cores, network latency and bandwidth, cache sizes, and local I/O throughput~\cite{cpuperformance, bader2021tarema}.
Accordingly, the same physical task will exhibit different runtimes on different nodes.
This opens the problem that schedulers need accurate task runtime estimates not only per task but actually per task-node pair.
Such information, however, is very often unavailable.

This paper presents Lotaru, a novel method addressing this problem through infrastructure profiling, local workflow executions on downsampled partitions of the entire input, and a Bayesian framework to transfer measured runtimes to specific task-node pairs.
It is intended for workflows for embarrassingly parallel problems where the same (sub-)workflows are executed over many inputs or intermediate results.
Examples of such problems are abundant in areas like remote sensing (sub workflows analyze many images in parallel~\cite{frantz2019force,berriman2004montage}), bioinformatics (sub workflows analyze many read sets in parallel~\cite{yates2021reproducible,garcia2020sarek}), or material science (sub workflows analyze many molecules in parallel~\cite{schaarschmidt2021workflow,stein2019progress}). In Figure~\ref{fig:execution_model}, the sub-workflow is executed once for each of the two input files.

In Lotaru's first step, the infrastructure profiler analyzes the performance characteristics of a local computer and the nodes in the cluster (called target nodes).
Subsequently, one of the foreseen input files is selected and downsampled or sliced into several smaller files, which next are used as input to run the workflow locally and to gather metrics.
The workflow is executed twice, once with all samples and once with a subset at reduced CPU speed to identify bottlenecks.
Next, Lotaru trains a Bayesian linear regression model for each task based on the collected runtimes.
This model is subsequently used to predict task runtimes for arbitrary input sizes.
Lastly, these predictions are adjusted to all target nodes using the initial  infrastructure profiling measurements.

We implemented Lotaru\footnote{github.com/CRC-FONDA/Lotaru} and compared it experimentally to three different baselines on a cluster of six different kinds of nodes using five real-life scientific workflows from the popular nf-core\footnote{nf-co.re} workflow repository~\cite{ewels2020nf} with different inputs.
Our experiments based on the popular workflow engine Nextflow~\cite{nextflow} show  that Lotaru significantly outperforms the baselines regarding prediction errors for homogeneous and for heterogeneous clusters.

Lotaru is designed as an online method, i.e., it does not depend on any historic information but performs all measurements and estimations before the start of a specific workflow execution.
Notably, this also allows for offline scenarios where the learned models are reused for future executions of the same workflow over different input data.
To this end, we also make available $9,317$ unique task execution traces from our experiments on six different machines\footnote{github.com/CRC-FONDA/Lotaru-traces}.



\section{Related Work}\label{sec:RELATED_WORK}
First, we cover scientific workflow management systems (SWMS) and scheduling strategies that could be used by such systems to show the need for runtime estimates.
We discuss approaches focusing on runtime prediction in general and finally focus on cross infrastructure task-runtime prediction.
Note that in the following, we focus mostly on the prediction of runtimes.
However, estimation methods for the usage of other resources, e.g. main memory or network bandwidth, often apply similar techniques~\cite{witt2019feedback, witt2019predictive}.

\subsection{Scientific Workflow Management Systems and Resource Manager}

SWMS like Pegasus~\cite{pegasus}, Saasfee~\cite{saasfee}, and Nextflow~\cite{nextflow} use workflow languages to define data analysis workflows in an abstract manner.
When executing a workflow on concrete inputs, they derive the physical workflow consisting of physical tasks and steer their execution with the help of a distributed resource manager, such as Slurm~\cite{slurm}, Kubernetes~\cite{kubernetes}, or YARN~\cite{vavilapalli2013apache}.
The resource manager coordinates the target infrastructure and acts as an intermediary between the SWMS and the cluster resources.
Current systems often do not perform any advanced form of scheduling because they do not have the capabilities to perform resource predictions.
Instead, SWMS send each ready-to-run task to the resource manager together with the user-defined requirements.
The resource manager in turn schedules them for execution in a FIFO or fair manner.
For example, several YARN distributions use a fair-like scheduler\footnote{docs.datafabric.hpe.com/62/AdministratorGuide/Job-Scheduling.html}\footnote{bdlabs.edureka.co/static/help/topics/admin\_fair\_scheduler.html}.

\subsection{Scheduling Workflow Tasks onto Heterogeneous Clusters}
\label{subsec:rw-scheduling}

Despite their missing uptake in real systems, the literature on more advanced scheduling algorithms is vast.
Scheduling of tasks onto heterogeneous infrastructures can be done in two ways, either statically or dynamically~\cite{dubey2018modified, wang2016hsip}.
Static scheduling heuristics like HEFT~\cite{heft} and HCPPEFT~\cite{dai2014synthesized} map all tasks to computing resources before the workflow execution.
Therefore, these approaches cannot adapt to infrastructure failures or changes in the physical workflow execution plan.
In contrast, dynamic scheduling approaches like P-HEFT~\cite{pheft} and FDWS~\cite{arabnejad2012fairness} map tasks to infrastructure components at runtime and are therefore more flexible.
In particular, they can also be applied when the set of physical tasks depends on intermediate results, in which case the complete physical workflow cannot be fixed before execution~\cite{bux2017hi}.
Both approaches have in common that they need at least comprehensive knowledge about execution times of all tasks on all available nodes.
However, these values are not available in advance but must be determined either by asking users for estimates~\cite{ilyushkin2018impact,feitelson2015workload, hirales2012multiple}, by analyzing historical traces~\cite{scheinert2021bellamy, will2021c3o, scheinert2021potential}, or by using some form of online learning~\cite{witt2019feedback, witt2019predictive}.
Lotaru aims to estimate the runtime for all task-node pairs in a cluster to enable the use of existing scheduling methods in real-world systems.
\subsection{Task Runtime Prediction Based on Historical Runtime Data}\label{subsec:approaches-for-runtime-prediction-based-on-historical-runtime-data}%
Runtime prediction for scientific workflow tasks based on historical data has been extensively researched.
Recent approaches use machine learning for this problem~\cite{da2013toward, nadeem2017modeling, da2015online, sadjadi2008modeling}.

\citet{nadeem2017modeling} propose the use of neural networks to predict workflow execution times on a grid.
Their learning model considers several types of information, like workflow structure, measured task resource requirements in historical traces, and input file sizes.
They further study which features influence the runtime the most and which ones can be omitted.

\citeauthor{da2013toward}~\cite{da2013toward, da2015online} estimate runtime, disk space, and memory consumption for tasks in scientific workflows.
They use monitoring tools and historical data to apply regression trees for actual predictions.
In a pre-processing step, they use density-based clustering to identify data subsets with high correlation.
If the data in the selected cluster is correlated, the authors estimate the expected resource usages based on the ratio in the cluster.
For uncorrelated data, the authors test a Gamma and a Normal distribution to generate an estimation value.
The estimations are updated at workflow runtime once new information becomes available.
As we use them as competitors for Loratu, we will describe them in more detail in Section~\ref{baselines}.

We call such approaches offline estimators, because they build their initial models on historical data prior to the actual workflow execution.
Note that offline estimators are generally not applicable for new workflows with new tasks, as for these, no historical traces for model learning are available.
In contrast, Lotaru is designed as an online method which is independent from any historical traces and can be applied out-of-the-box for any kind of workflow on any kind of cluster.

\subsection{Cross-Infrastructure Task-Runtime Prediction }
\label{subsec:cross-infrastructure}

Because real clusters are often comprised of nodes with heterogeneous capabilities, all methods that learn a model from past task executions (offline or online) have to consider the challenge that their models must generalize to different nodes.
This is particularly important when running a workflow on a completely different cluster, i.e., in a cross-infrastructure setting.

\citet{pham2017predicting} use a two-stage prediction approach to estimate the task execution time in cloud environments.
As prediction parameters, they consider input parameters of the workflow, VM specifications, and runtime parameters like memory usage and read/write operations.
In the first stage, the authors derive the runtime parameters for the execution on the target VM.
The second stage uses the output data from the first stage together with workflow input data and the VM specifications to learn a regression model that predicts the execution time of a given task on a target VM.

\citet{hilman2018task} apply an online incremental learning approach using long short-term memory networks (LSTMs) to predict task runtimes.
As input features, they consider task characteristics, like the name of the executable, the input data, VM details, and the submission time to encounter for daily trends in resource usage.
These features are extended through historical time-series data for CPU usage, memory usage, and disk activities to conduct their incremental prediction method.

\begin{figure*}
    \includegraphics[width=\textwidth]{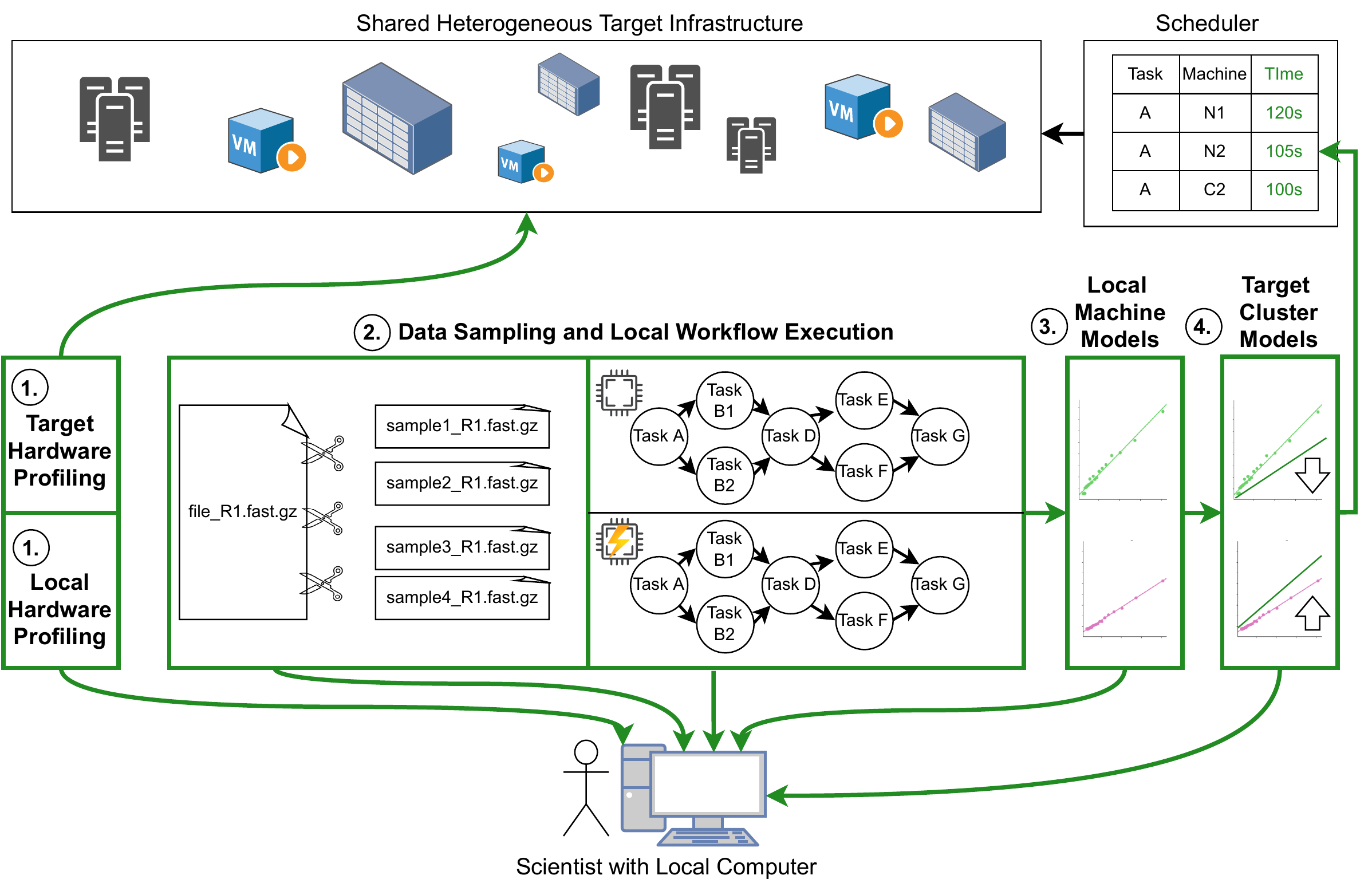}
    \caption{The process of Lotaru's local runtime prediction approach.} \label{fig:arch}
\end{figure*}

\citet{matsunaga2010use} include system-specific attributes like CPU architecture and the size and speed of the memory to derive more accurate prediction models.
Their study compares several machine learning approaches for two well-known bioinformatics applications.

Lotaru does not include explicit hardware characteristics as features for the runtime prediction as they are often difficult to correlate with the runtime of real tasks.
Instead, we apply microbenchmarks for obtaining such characteristics in an implicit manner.
Furthermore, many of the recent approaches use machine learning methods like neuronal networks or k-nearest neighbors, which are known to require large training data sets to perform well.
In contrast, Lotaru uses a Bayesian Linear Regression that already works with few training points and that is also capable of deriving  uncertainty estimates for its predictions~\cite{mcneish2016using, lee2004evaluation}.

\section{Lotaru Approach}\label{sec:APPROACH}
Lotaru is a novel method that aims to predict workflow task runtimes for all nodes in heterogeneous clusters at workflow start-up time and without relying on historical data.
Figure~\ref{fig:arch} provides an overview of our approach.
In phase \textcircled{\raisebox{-0.9pt}{1}}, Lotaru uses microbenchmarks to gather performance insights about the target infrastructure and the local developer machine.
In the second phase \textcircled{\raisebox{-0.9pt}{2}}, Lotaru selects one of the data inputs and downsamples it to create several small workflow input partitions.
Then, the workflow is locally executed two times with the created partitions.
For the second execution, only a few of these partitions are used, and the CPU speed is slightly reduced to identify CPU-bound tasks.
In the third phase \textcircled{\raisebox{-0.9pt}{3}}, Lotaru uses the collected data points to create a Bayesian linear regression model which predicts the runtime.
In the last phase \textcircled{\raisebox{-0.9pt}{4}}, the values from the benchmarks are used to adjust the prediction results for each node in the target infrastructure. 
The scheduler can then consider the prediction results for all task-node pairs to create a scheduling plan.

One assumption behind Lotaru is that the input consists of multiple input files, which are, at least partly, processed separately.
We rely on this property to create local workflows with small inputs for fast measurements of first characteristics. 
Furthermore, in this work, we assume that Lotaru can downsample or slice an input file further to create a diverse set of local workflow inputs, allowing us to learn the relationship between input size and task runtimes.
However, in applications where such a downsampling is not possible, Lotaru could also use different subsets of the input files at the price of a somewhat longer training phase.
\subsection*{\textcircled{\raisebox{-0.9pt}{1}} Local and Target Infrastructure Profiling}
\label{subsec:approach-profiling}

We expect the local developer machine to differ from the target cluster and the target nodes themselves to be heterogeneous.
To measure the differences, we conduct a short profiling phase to gather detailed infrastructure characteristics. 
Therefore, Lotaru analyzes all target nodes' dynamic performance characteristics like CPU speeds, memory speed, and random and sequential I/O.
For this, we use microbenchmarks, which can be executed in parallel and take very short time, typically less than a minute, for each node. 
This step could be rerun automatically whenever a cluster's resource manager detects hardware changes.

\subsection*{\textcircled{\raisebox{-0.9pt}{2}} Data Sampling and Local Workflow Execution}
\label{subsec:approach-sampling}

In the next step, Lotaru trains a Bayesian regression model to predict task runtimes based on input size. 
To this end, it picks one of the input files and downsamples it further to obtain task runtimes for diverse yet small (and hence fast) inputs as input for the learner.
For image data used in remote sensing or astronomic workflows, this means dividing a single image into smaller ones keeping the resolution, or decreasing the resolution while leaving the image section the same.
In genomics, downsampling means splitting one of the many samples with millions of short sequence reads into multiple smaller partitions.

Next, Lotaru measures local runtimes over multiple different partitions.
While using a large set of such partitions covering a large range of data sizes tend to improve the accuracy of the prediction model, fewer and smaller partitions can be executed faster and lead to quicker but mostly more imprecise runtime estimates.

If, for instance, the data sampling process described before created five partitions, Lotaru runs the workflow with these five input files.
This step delivers monitoring data for each task-partition pair but gives no direct insights whether, i.e., a task is CPU-intense.
To identify which hardware resource the task mostly depends on, we decrease the CPU frequency of the local machine by 20\% and run the workflow again with the five created partitions.
Thereby, we expect CPU-intense tasks to take around 25\% longer.

The sizes of the samples obviously are an important hyperparameter of Lotaru whose effect will be studied in Section~\ref{sec::ImpactOfDownsampling}.

\subsection*{\textcircled{\raisebox{-0.9pt}{3}} Local Prediction Model Training}
\label{subsec:approach-model}

\begin{table*}[]
\caption{Example of a model adjustment for a single task prediction on two target nodes.}
\begin{tabular}{|c|r|r|r|r|r|r|r|}
\hline
Machine & CPU                         & RAM                           & I/O                         & Weight $w$ & Bayesian Prediction & Factor                        & Final Runtime Prediction               \\ \hline
Local   & \cellcolor[HTML]{C0C0C0}500 & \cellcolor[HTML]{C0C0C0}20,000 & \cellcolor[HTML]{C0C0C0}500 & 0.80             & 100.00s                 & 1.00                             & 100.00s                          \\ \hline
N1      & \cellcolor[HTML]{C0C0C0}400 & \cellcolor[HTML]{C0C0C0}18,000 & \cellcolor[HTML]{C0C0C0}300 & -                & -                   & \cellcolor[HTML]{FFC702}1.33  & \cellcolor[HTML]{FFC702}133.00s  \\ \hline
N2      & \cellcolor[HTML]{C0C0C0}520 & \cellcolor[HTML]{C0C0C0}20,000 & \cellcolor[HTML]{C0C0C0}500 & -                & -                   & \cellcolor[HTML]{FFC702}0.96 & \cellcolor[HTML]{FFC702}96.00s \\ \hline
\end{tabular}
\label{tab:example-runtime-estimate}
\end{table*}

Most existing approaches use the file size on disk as the input or part of the input vector for their predictions or statistical models.
We argue that Lotaru should use the uncompressed input data size for compressed files, which scientific workflows frequently use due to the large amount of data.
For example, in bioinformatics workflows, the de facto standard file format for storing biological sequences is fastq which is compressed with Gzip.
Gzip can compress larger files efficiently, especially when dealing with repetitive data, leading to a non-linear file size increase.
For example, the file example.fastq.gz\footnote{s3://nf-core-awsmegatests/eager/ENA\_Data\_Fish/ERR1943601\_1.fastq.gz} has a size of $\sim$2,014~MB and contains 40,517,845 sequences.
Splitting the file into two files with the same number of sequences leads to two files with a size of $\sim$1,274~MB each, an increase of 26.46\%.
Accordingly, Lotaru should not use the compressed file size to predict the runtime.

Using the uncompressed data size and a task's runtime, Lotaru checks for a linear correlation between both.
For this, we use the Pearson correlation coefficient, which is defined as follows:

\begin{equation}
\begin{aligned}
    p = \frac{\sum (x_i - \bar{x}) (y_i - \bar{y})}{\sqrt{\sum (x_i - \bar{x})^2 \sum(y_i -\bar{y})^2}}
\end{aligned}
\end{equation}
\noindent
where the x values are the actual uncompressed task input data size and y values are the runtime.
We define a correlation as significant if $p$ is greater than 0.8.
Lotaru uses a Bayesian linear regression to predict the runtime if $p$ is significant.
Otherwise, we estimate the median runtime as the expected runtime independent of the concrete input size.

One of the main advantages of using the Bayesian approach is that Lotaru can train it on a small training data set~\cite{mcneish2016using,lee2004evaluation}, which is especially useful since the local profiling only delivers a few training points for each task.
Additionally, while Lotaru can predict the most probable runtime, the Bayesian approach also yields an uncertainty value for this prediction.
For example, Lotaru estimates that task A is expected to run 120 seconds.
However, it also gives a lower and upper uncertainty at different confidence levels to express that the point estimate probably is not accurate.

\begin{figure}
\centering
    \includegraphics[width=0.45\textwidth,trim={0mm 15mm 2mm 0mm}]{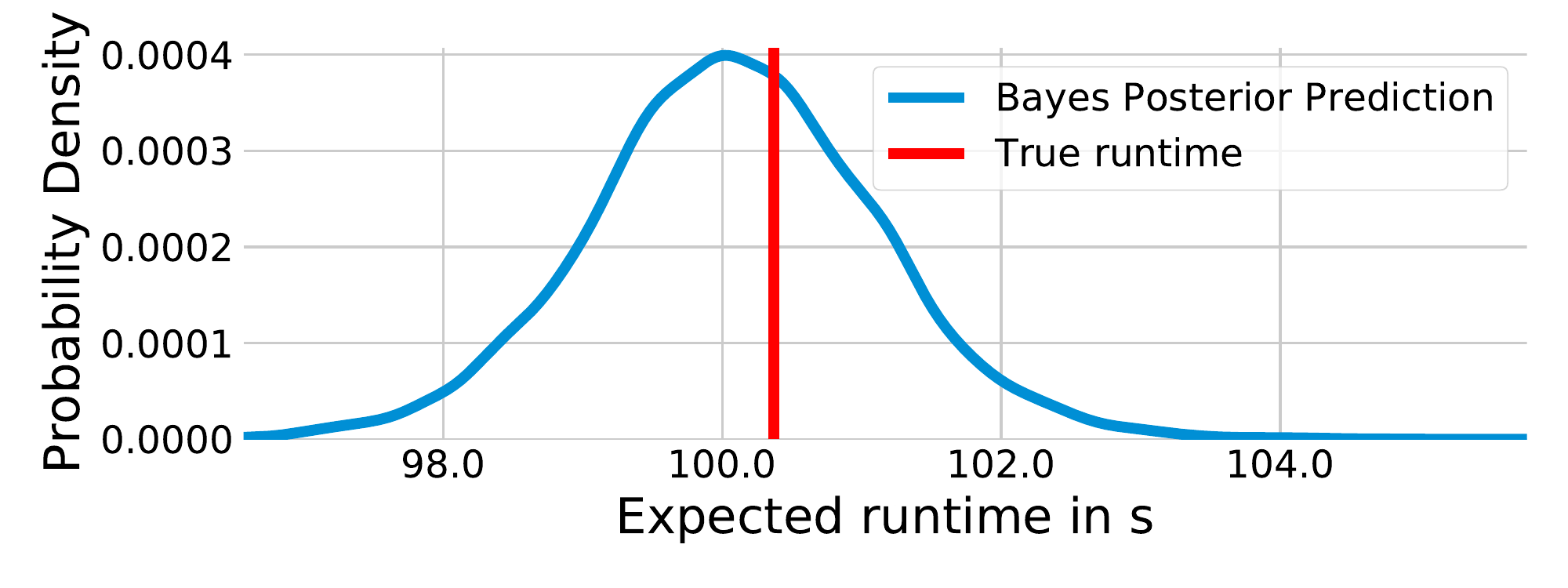}
    \caption{Posterior Prediction for the task FASTQC with an uncompressed input size of $\sim$4,585~MB.} 
    \label{fig:bayes_posterior}
\end{figure}
As illustration, Figure~\ref{fig:bayes_posterior} shows the prediction for the task FASTQC in the Chipseq-1 workflow with an input data size of $\sim$4,585~MB (see Section~\ref{subsec:eval-data} for details on the data and the workflow).
The prediction error for the estimated mean value is around 0.30\%.
The value is in a confidence interval of 23.23\% uncertainty.
With a confidence of 50\% uncertainty, we would consider the runtime to be between 99.4s and 100.7s.
The scheduler can consider this uncertainty and plan with it, which would not possible for frequentist approaches.

In contrast to these classical frequentist approaches, we try to find a posterior distribution for our model parameters.
Specifically, our model computes a posterior distribution depending on the input values as shown in Formula~\ref{formula:bayes_1}, we assume that $\epsilon_i$ and $y_i$ are normally distributed with $\epsilon_i \sim N(0, \sigma^2) $ and $ y_i \sim N(b^T x_i;\sigma^2)$

\begin{equation}
\begin{aligned}
\label{formula:bayes_1}
    y_i = x_i^T b + \epsilon_i;
\end{aligned}
\end{equation}

Our Bayesian approach now tries to maximize the posterior

\begin{equation}
\begin{aligned}
    \max_{b} P(b|y_i);
\end{aligned}
\end{equation}

Through the rules of Bayes and a short equation transformation, we get to

\begin{equation}
\begin{aligned}
    \max_{b} P(y_i|b) P(b);
\end{aligned}
\end{equation}

where the first term, $P(y_i|b)$, the likelihood, can be computed with our previous assumption about the distribution of $y_i$. For the second term, the posterior, $P(b)$, we have to assume a distribution.
We decided to set the prior to a Gaussian distribution, which results in an L2-regressor for our Bayesian regression.

In our prediction model, $x_i$ is a scalar, i.e., the uncompressed task data input size, and $y_i$ is the runtime.

\subsection*{\textcircled{\raisebox{-0.9pt}{4}} Model Adjustment for Target Infrastructure}
\label{subsec:approach-targetCLuster}

With the Bayesian Linear Regression model created, Lotaru can now predict the tasks' runtime for nodes with the same hardware as the developer machine.
However, we aim to estimate runtimes for all different kinds of nodes in the cluster.
Therefore, we take the monitoring data from the infrastructure profiling and the local workflow runs.
For each abstract task, we compare the runtimes for the task-sample pairs which occurred in the two local workflow executions, the normal one and the one with reduced CPU speed.
Lotaru determines the deviation of a task's pair's runtime as $dev = \frac{time_{new} - time_{old}}{time_{old}}$, where $old$ refers to the normal execution and $new$ to the execution with reduced CPU speed. 

Since each task at least has to read the input file and to write the output file, I/O capabilities are essential for every task and are, therefore, included in the following adjustment step.
Another important factor is CPU speed, whereas in prior experiments we found that memory speed only has a negligible impact.
Thus, we decided to exclude the memory speed from our adjustment step.
Accordingly, we define the impact of the CPU and the I/O on a task's execution time through the following weighting:
\begin{equation}
\begin{aligned}
\label{formula:weighting}
    w = max \left( 0; min \left( 1; \frac{median_{dev}}{(freq_{old} / freq_{new}) -1} \right)\right)
\end{aligned}
\end{equation}

Then, Lotaru sets the runtime factor for each task as following:

\begin{equation}
\begin{aligned}
  f_t = w * \frac{cpu_{local}}{cpu_{target}} +
        (1 - w) * \frac{io_{local}}{io_{target}}
\end{aligned}
\label{equation::machineFactor}
\end{equation}

Table \ref{tab:example-runtime-estimate} gives an example for a single task and a prediction on two different target nodes, namely N1 and N2.
The values with the grey background are from the profiling step, and the CPU-weight $w$ for the factor is 0.8. 
For one out of the various inputs, e.g., with a size of 10GB, a runtime of 100 seconds is predicted on the local machine with the Bayesian regression model.
We can now adjust the estimated runtime to the target nodes by calculating the factor with the monitoring values.
The results are depicted in yellow.

\section{Experimental Setup}\label{sec:EXPSetup}
This section presents our prototype implementation and the environment where we run our experiments.
The source code to reproduce the evaluation is available online\footnote{github.com/CRC-FONDA/Lotaru}.

\subsection{Prototype Implementation}
\label{subsec:eval-prototype}

\paragraph{\textbf{Infrastructure Profiler:}}

\begin{table*}[]
\centering
\caption{The results from applying the infrastructure profiling on the six different nodes.}
\begin{tabular}{|c|c|r|c|r|r|r|r|r|}
\hline
Machine & \# CPUs & Memory & Storage & CPU events/s & LINPACK & RAM score & read IOPs & write IOPS \\ \hline
Local   & 8    & 16 GB     & HDD     & 458  & 3,959,800    & 18,700     & 414       & 415        \\ \hline
A1   & 2 x 4    & 32 GB     & HDD     & 223  & -    & 11,000     & 306       & 301        \\ \hline
A2   & 2 x 4    & 32 GB     & HDD     & 223  & -    & 11,000     & 341       & 336        \\ \hline
N1      & 8    & 16 GB    & HDD     & 369   & 3,620,426   & 13,400     & 481       & 483        \\ \hline
N2      & 8    & 16 GB    & HDD     & 468   & 4,045,289   & 17,000     & 481       & 483        \\ \hline
C2      & 8    & 32 GB    & HDD     & 523   & 4,602,096   & 18,900     & 481       & 483        \\ \hline
\end{tabular}
\label{tab:benchmakred_machines}
\end{table*}

\begin{table*}[]
\centering
\caption{The workflows used in our experiments with their input data and key characteristics.}
\begin{tabular}{|c|r|r|r|r|r|}
\hline
Workflow                   & \# Abstract Task Definitions                                                            & Sample & Size & Uncompressed Size & Workflow Runtime One Input \\ \hline
\multirow{2}{*}{Eager}     & \multirow{2}{*}{13}                         & 1         & 1.52 GB        & 8.33 GB    &   148 min  \\ \cline{3-6} 
                           &                                                                        & 2         & 4.34 GB       & 25.71 GB  &   211 min  \\ \hline
\multirow{2}{*}{Methylseq} & \multirow{2}{*}{8}                  & 1         & 3.61 GB       & 17.03 GB  &   90 min  \\ \cline{3-6} 
                           &                                                                        & 2         & 4.75 GB       & 22.50 GB  &   117 min  \\ \hline
\multirow{2}{*}{Chipseq}   & \multirow{2}{*}{14}                          & 1         & 1.33 GB       & 4.81 GB   &   140 min  \\ \cline{3-6} 
                           &                                                                        & 2         & 8.71 GB       & 32.98 GB  &   948 min \\ \hline
\multirow{2}{*}{Atacseq}   & \multirow{2}{*}{14}                          & 1         & 3.26 GB       & 14.09 GB   &   184 min  \\ \cline{3-6} 
                           &                                                                        & 2         & 2.40 GB       & 11.81 GB  &   104 min \\ \hline
\multirow{2}{*}{Bacass}   & \multirow{2}{*}{5}                          & 1         & 1.23 GB       & 3.64 GB   &   237 min  \\ \cline{3-6} 
                           &                                                                        & 2         & 1.45 GB       & 4.35 GB  &   253 min \\ \hline
\end{tabular}
\label{tab:workflows}
\end{table*}

The infrastructure profiler uses \textit{sysbench}\footnote{github.com/akopytov/sysbench} as a first measuring tool and \textit{LINPACK}~\cite{dongarra2003linpack} as a second microbenchmark since both measure different CPU characteristics and are well-known.
With \textit{sysbench}, we run a benchmark that verifies prime numbers with a limit of ten seconds and a maximum verification prime number of 20,000.
Additionally, \textit{LINPACK} measures the FLoating Point Operations Per Second (FLOPS) using a default array size of $200 x 200$.
The memory speed is tested through \textit{sysbench}, setting the block size buffer to one megabyte and the total memory size to 100 gigabytes.

Since we run \textit{sysbench} on computers that differ in the number of CPU cores, we decided to always set the number of benchmarked CPU threads to one.
Therefore, we differentiate between a node's overall and single-core performance because the resource manager assigns a fixed number of cores to workflow tasks.
Therewith, we have a better comparability, and Lotaru avoids that nodes with more but weaker CPU cores score higher.
Otherwise, for example, a node with four very powerful cores would score less than a node with eight slower cores.

Lotaru tests the I/O performance of the local machine and target cluster by using \textit{fio}\footnote{github.com/axboe/fio}.
We conduct a benchmark only for sequential read-write and avoid measuring random read-write characteristics since this access pattern is scarce in scientific data analysis.
Typically big input files are read sequentially.
Further, we omit memory use and the page cache, so these components do not influence the I/O measurements.

Nowadays, hardware is tailored to achieve many points in popular benchmarks like \textit{sysbench} and \textit{fio}.
Note, however, that our goal is not to benchmark the actual performance of the hardware but to compare the performance of different nodes for adapting runtime estimates.

\paragraph{\textbf{Data Sampling and Local Runtime Prediction:}}
Lotaru relies on downsampling of workflow input data to obtain measurements for task executions quickly.
Such downsampling obviously must take the actual nature of the data into account and therefore must be implemented specifically for different input data.
As all our evaluation workflows run on genome sequencing data, we implemented downsampling for genome data in the fastq format using the open-source software \textit{fastqsplitter}\footnote{github.com/LUMC/fastqsplitter} to split the inputs in partitions.
Note that Lotaru features an interface to support downsampling or slicing files in arbitrary domains.

For managing the CPU frequency, we use the userspace tool \textit{cpupower}\footnote{linux.die.net/man/1/cpupower}.
As a workflow management system, we choose \textit{Next\-flow}~\cite{nextflow}, which gathers task runtime metrics already by default.
We extended the monitoring interface of \textit{Nextflow} to collect additional data, like compressed and uncompressed input size of tasks and the overall workflow input size.

\subsection{Cluster Setup and Evaluation Workflows}
\label{subsec:eval-data}
We evaluate our approach on six different machines: a local machine, two machines from a heterogeneous commodity cluster, and three virtual machines in the Google Cloud Platform (GCP). Specifications can be found in Table~\ref{tab:benchmakred_machines} together with results of our microbenchmarks. The local machine consists of an Intel Xeon E3-1230 V2 CPU (four cores, eight threads, 3.30 GHz base frequency), 16GB memory, and an HDD.
The two machines from the commodity cluster, A1 and A2, have two Intel Xeon X5355 2.66 GHz each, 32GB of memory and different hard drives.
Since the \textit{LINPACK} benchmark failed on A1 and A2, due to the age of the machines, the values are not included in the table and only the \textit{sysbench} score is used for the factor.
We use N1, N2, and C2 instances as (heterogeneous) nodes in the cluster.
While the N1 machines are based on Intel Broadwell with a base clock of 2.0 GHz, the N2 machines use Intel Cascade Lake CPUs with a base clock of 2.8 GHz.
The C2 machines are compute-optimized and based on Intel Cascade Lake with a turbo clock of up to 3.8 GHz\footnote{cloud.google.com/compute/docs/machine-types}.

We selected five real-world bioinformatics workflows from the nf-core repository\footnote{github.com/nf-core} and ran each of them with two different data sets to evaluate our approach.
Table~\ref{tab:workflows} gives an overview of these workflows and data inputs.
Each of the five workflows performs different types of sequence analysis: The Eager workflow~\cite{yates2021reproducible} analyzes ancient genomic data, the Chipseq workflow\footnote{github.com/nf-core/chipseq} is used to analyze Chromatin ImmunopreciPitation sequencing (ChIP-seq) data, and the Methylseq workflow\footnote{github.com/nf-core/methylseq} is used for analyzing Bisulfite sequencing data in epigenomics.
Atacseq\footnote{github.com/nf-core/atacseq} analyzes ATAC-sequencing data and Bacass\footnote{github.com/nf-core/bacass} is a workflow for simple bacterial assembly and annotation.
The workflow runtimes in Table~\ref{tab:workflows} were obtained by running the workflows with one of various data inputs on the local machine. 
Note, typically, workflows run hundreds or thousands of inputs, resulting in much higher workflow runtimes for real inputs, even on large-scale clusters.

\subsection{Baselines}\label{baselines}

We compare the accuracy of Lotaru's runtime predictions with three baselines: a Naive Approach (NA), Online-M~\cite{da2013toward}, and Online-P~\cite{da2015online}.

The Naive Approach estimates the ratio $r_t = \frac{run_q}{d_q}$ for each training data tuple $q$ (uncompressed input size $d_q$, runtime $run_q$) and takes the mean $\bar{r_t}$ for task $t$ over these ratios.
It then uses this mean ratio to predict the runtime of a task $t$ with uncompressed target input size $d_t$, using $\bar{r_t} * d_t$. 
Online-P and Online-M use density-based clustering to identify high-density areas.
Then, a cluster is determined according to the I/O read value of the estimated task.
Since the clustering is not possible with the sparse data from the local executions, we take the data point closest to the task being estimated.
Then, a pearson correlation between all input and output parameters is calculated.
If the data correlates, the ratio between output and input parameter is computed and used for the prediction.
If the data is uncorrelated, Online-M directly estimates the mean, while Online-P first tries to sample from a Normal or Gamma distribution.
Both approaches monitor the workflow execution and can update the estimates as more information becomes available.
However, this is not implemented since we focus on the out-of-the-box prediction accuracy without historical data.

\section{Evaluation Results}\label{sec:Eval}

We run three types of experiments. 
First, we test different downsampling combinations and sizes to evaluate their impact on Lotaru's prediction errors.
In the second experimental scenario, we train the task models on a similar machine as the target environment to get an unbiased view of the prediction models' capabilities.
The third scenario evaluates the setup in the heterogeneous target cluster.
Here, we have to adjust the runtime predictions from our local machine to the different nodes in the target cluster using the adjustment factor.

For our evaluation, we introduce the median prediction error (MPE). This metric is calculated for every workflow and aggregates the prediction error of the tasks inside the workflow.
To this end, we compute the prediction error for a single task as:

\begin{equation}
\begin{aligned}
    err_t = abs\left( \frac{predicted\_runtime - actual\_runtime}{actual\_runtime} \right), 
\end{aligned}
\end{equation}.

\begin{figure}
\centering
\begin{subfigure}{.23\textwidth}
   \centering
  \includegraphics[width=1\textwidth, trim={3mm 5mm 20.2mm 0mm}]{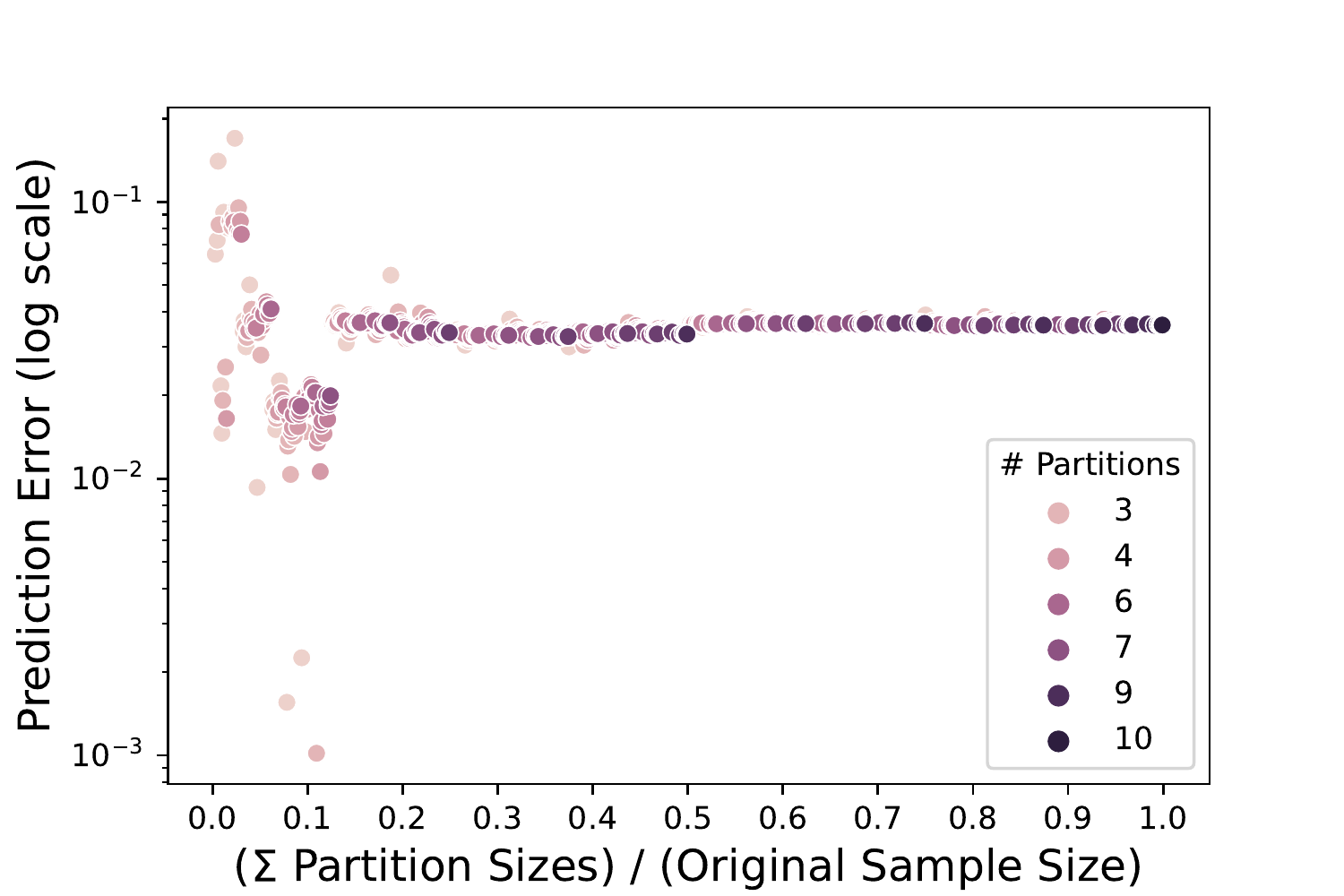}
  \caption{BWA task }
  \label{fig:sub-first}
\end{subfigure}
\begin{subfigure}{.23\textwidth}
  \centering
  \includegraphics[width=1\textwidth,trim={3mm 5mm 20.2mm 0mm}]{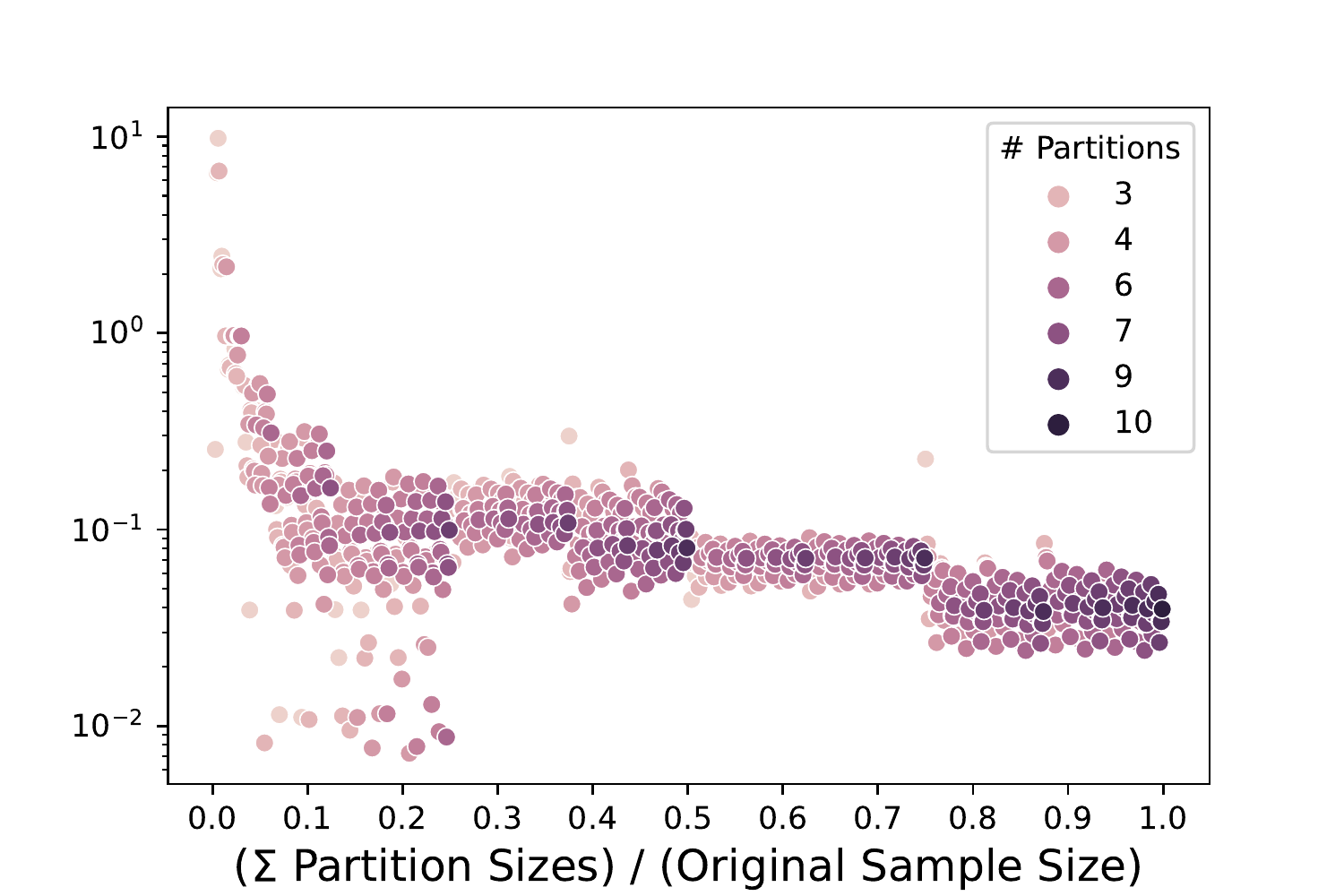}
  \caption{Mark duplicates task}
  \label{fig:sub-second}
\end{subfigure}
\begin{subfigure}{.23\textwidth}
   \centering
  \includegraphics[width=1\textwidth, trim={3mm 5mm 20.2mm 0mm}]{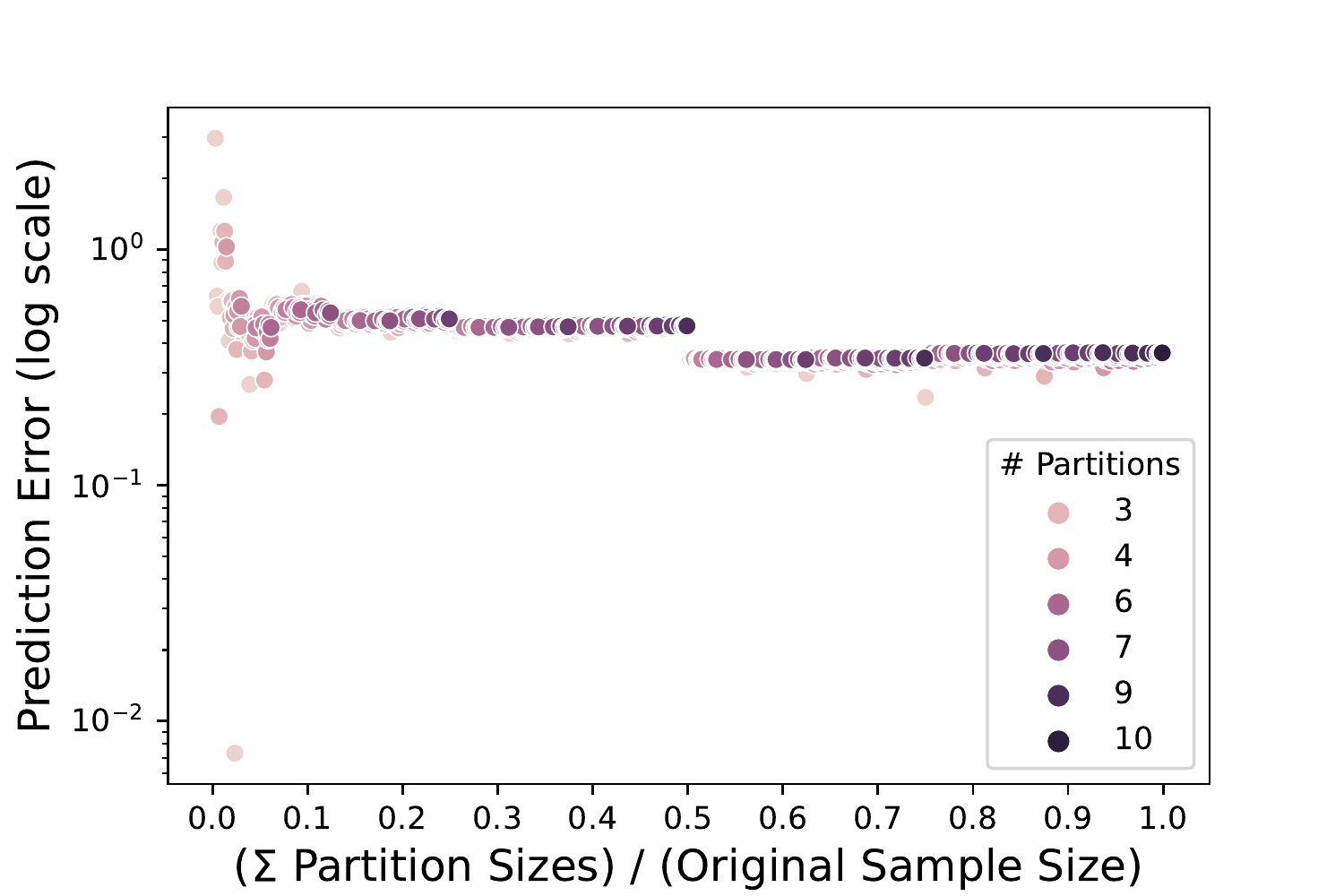}
  \caption{Genotyping task}
  \label{fig:sub-third}
\end{subfigure}
\begin{subfigure}{.23\textwidth}
  \centering
  \includegraphics[width=1\textwidth,, trim={3mm 5mm 20.2mm 0mm}]{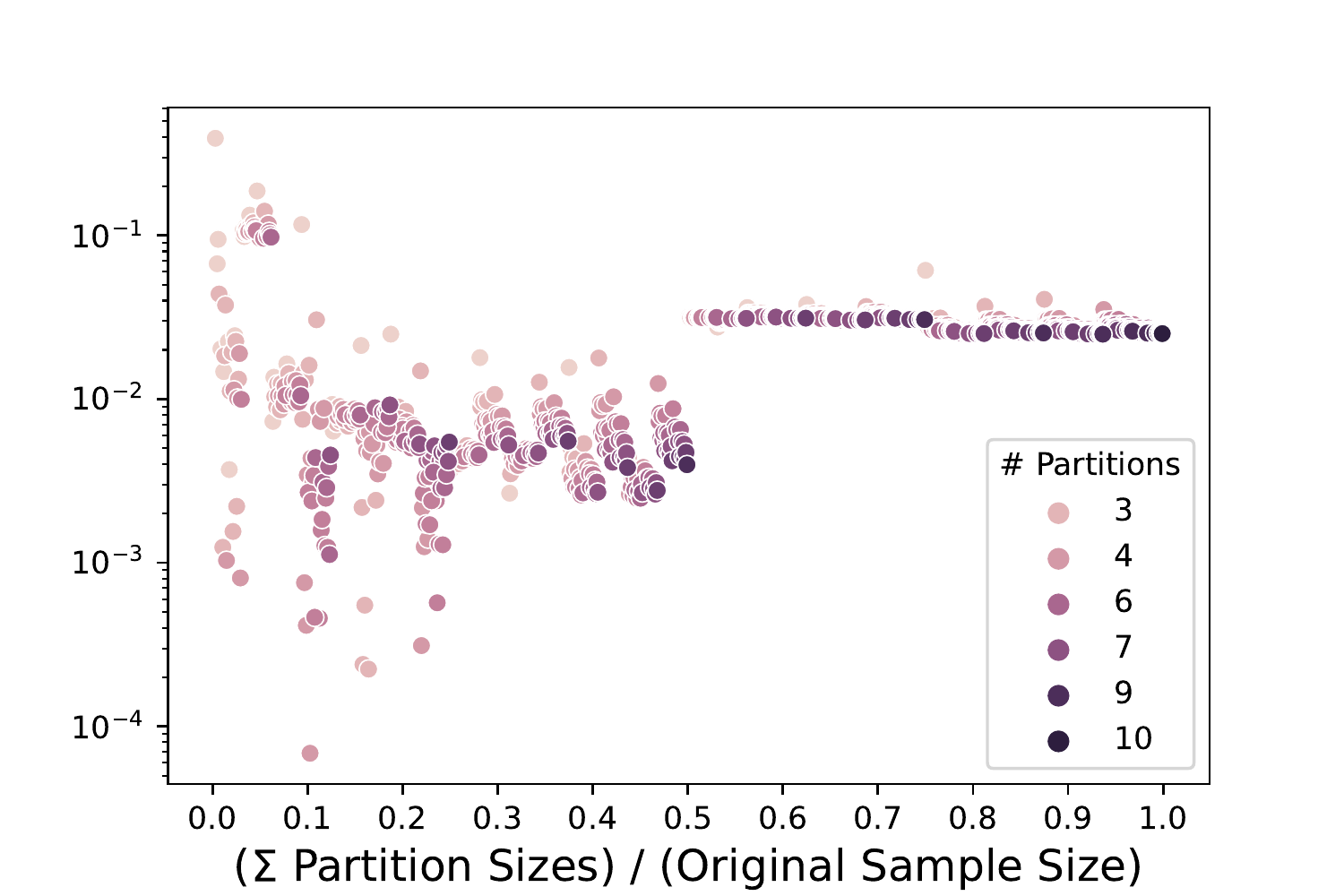}
  \caption{Adapter removal task}
  \label{fig:sub-fourth}
\end{subfigure}
\begin{subfigure}{.23\textwidth}
   \centering
  \includegraphics[width=1\textwidth, trim={3mm 5mm 20.2mm 0mm}]{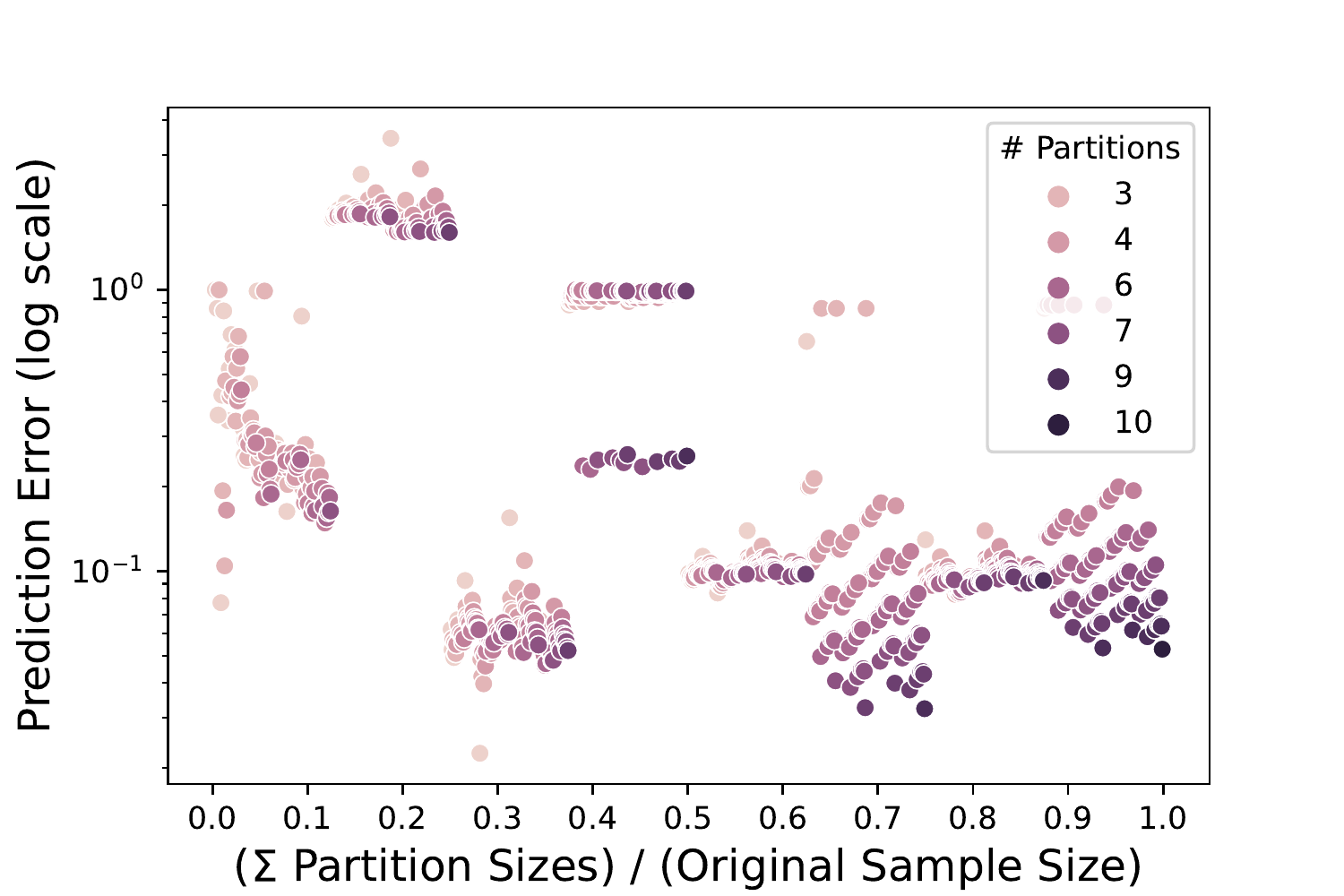}
  \caption{Samtools task}
  \label{fig:sub-fith}
\end{subfigure}
\begin{subfigure}{.23\textwidth}
  \centering
  \includegraphics[width=1\textwidth,, trim={3mm 5mm 20.2mm 0mm}]{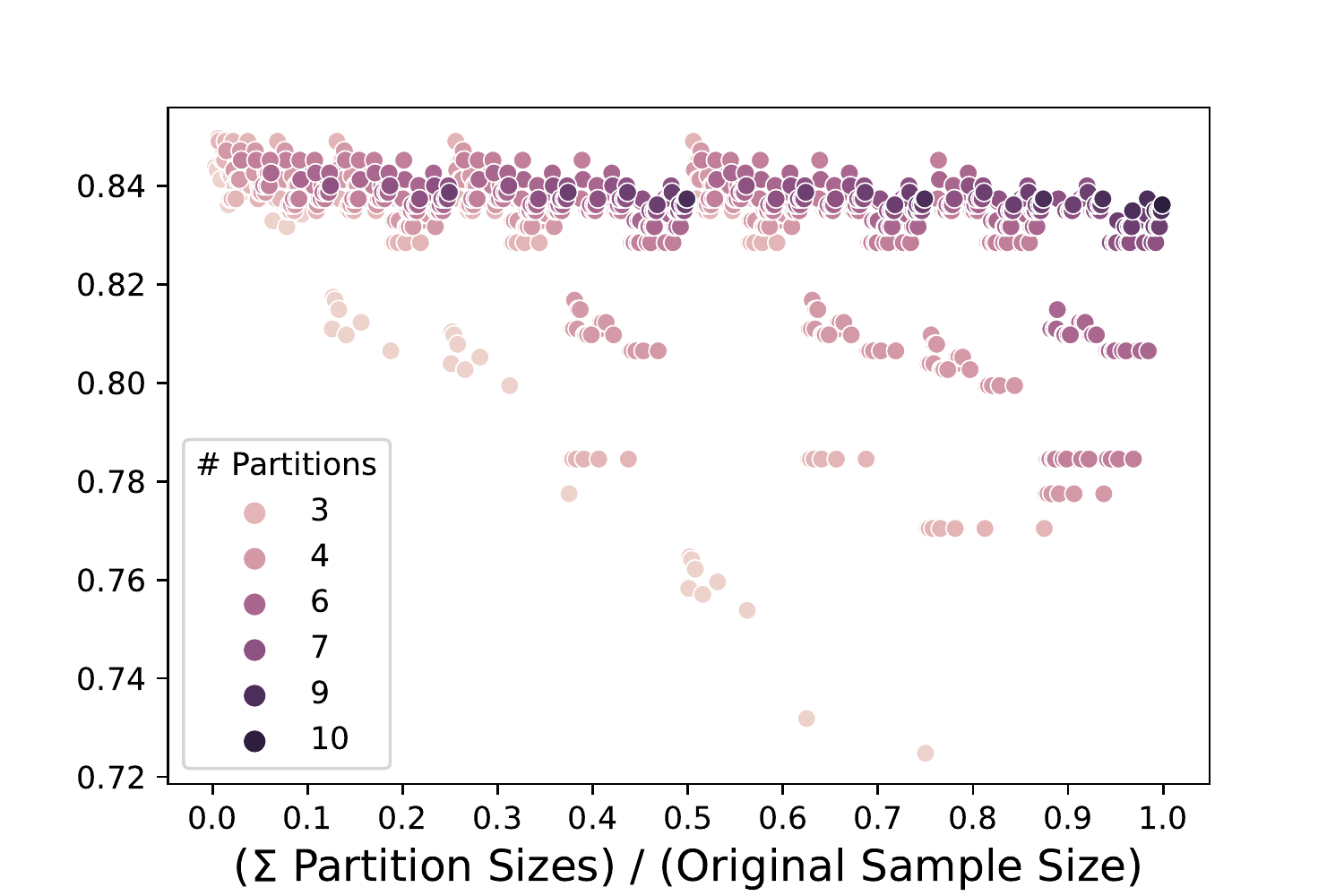}
  \caption{Bcftools task}
  \label{fig:sub-sixth}
\end{subfigure}
\caption{Relationship between the number and the cumulated size of the downsampled partitions and the prediction error for Lotaru and various tasks from the Eager-1 workflow.}
\label{fig:impactSampleSize}
\end{figure}%

\subsection{Impact of the Downsampling on Prediction Accuracy}\label{sec::ImpactOfDownsampling}
Lotaru takes one of the many workflow inputs and downsamples or slices this input for local workflow execution to gather training data.
This is a crucial step in our approach since the number and sizes of the chosen partitions highly influence the prediction error and the local workflow runtime.

Therefore, we first want to evaluate how many samples Lotaru should create from one original input and which sizes in relation to it are necessary to achieve good prediction results.
The experiment results can be generalized for genomic workflows.
Other domains, such as remote-sensing or material science, need a separate analysis.
Consequently, we designed our experiment the following:

\noindent{}For each pair of workflow and input, we cut one of the original input files with a size of $X$ into ten partitions for the Eager, Methylseq, Atacseq, and Bacass workflow and 16 partitions for the Chipseq workflow.
The size of the first partition, $s_1$, is set to $s_1 = \frac{X}{2}$, and $s_{n} = \frac{s_{n-1}}{2}$, so that $s_1$ has half, $s_2$ a quarter of the original size, and so on.
Then, we apply our prediction model to all possible partition combinations.
Therefore, ten possible input partitions, result in $\sum_{k=2}^{10} \frac{10!}{k!(10-k)!} = 1,013$ combinations for each task and prediction method.

Due to the high number of evaluated workflows with two different datasets used for each workflow, we decided to highlight the Eager-1 workflow as the example workflow but made all traces available online\footnote{github.com/CRC-FONDA/Lotaru-traces}.
Therefore, Figure~\ref{fig:impactSampleSize} shows Lotaru's prediction error for six representative tasks from the Eager-1 workflow, depending on the number of the downsampled partitions and their cumulative input size.
First, one can see that if the cumulative input size is below 10\% of the original input size, the predictions yield a high variance which is reflected in the prediction error.
Further, once an input combination reaches this threshold, the number of partitions does not affect prediction error significantly, as long as there are at least three partitions.

Out of the 13 abstract tasks from the Eager workflow, eleven tasks follow this pattern, the samtools tasks shows no relation, Figure~\ref{fig:sub-fith}, and the bcftools task is predicted according to the median, Figure~\ref{fig:sub-sixth}.
Additionally, our observations show that increasing the cumulative size can further reduce the prediction error but also increase the local runtime.
Examples for this are the tasks from the Figures~\ref{fig:sub-second} and ~\ref{fig:sub-third}, while the other figures do not follow this pattern.
The results from our other workflow executions underline our observations.

At the same time, the cumulative input data size for the local workflow execution determines the runtime on the developer's computer.
Most tasks have a linear relationship between input data size $s$ and runtime $r$.
Thus, we can extrapolate the runtime $r_s$ from one original input sample.

\begin{figure}
\centering
\begin{subfigure}{.2\textwidth}
   \centering
  \includegraphics[width=1\textwidth,trim={10mm 5mm 25.2mm 0mm}]{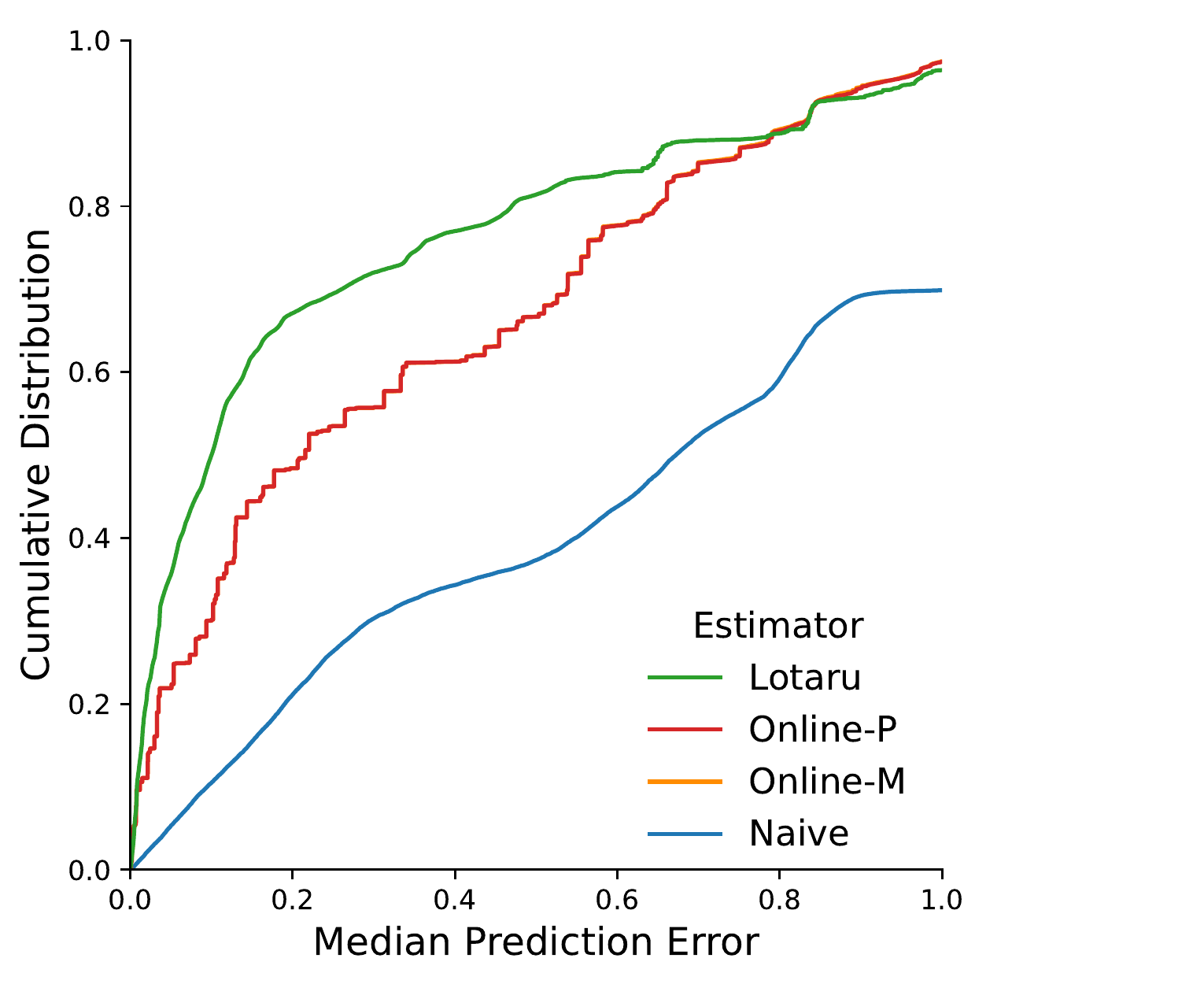}
  \caption{Prediction Errors \\ Eager-1 }
  \label{fig:cumu-sub-first}
\end{subfigure}
\begin{subfigure}{.2\textwidth}
  \centering
  \includegraphics[width=1\textwidth,trim={10mm 5mm 25.2mm 0mm}]{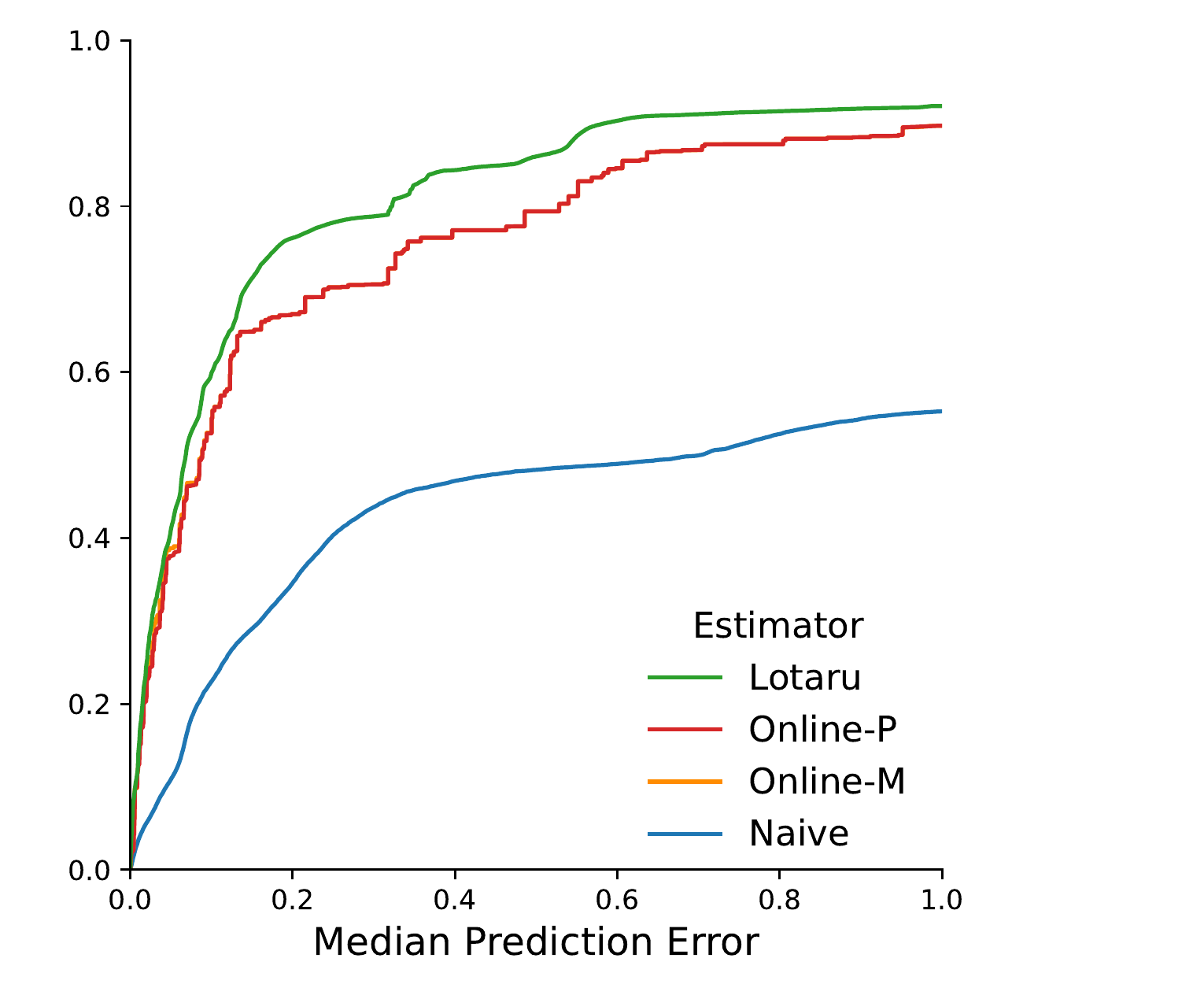}
  \caption{Prediction Errors \\ \centering Atacseq-1 }
  \label{fig:cumu-sub-second}
\end{subfigure}
\caption{Cumulative Distribution of the prediction error from the four approaches for two workflows. Red line similar to the orange line.}
\label{fig:cumDistr}
\end{figure}%

As examples, Figure~\ref{fig:cumDistr} shows the cumulative distribution of the prediction error for all tasks of the complete Eager-1 and Atacseq-1 workflow, and compares Lotaru to the three other approaches.
The Atacseq-1 workflow is a setup where the baselines Online-P and Online-M show a similar error distribution for around 38\% of all combinations compared to Lotaru.
The plot line for Online-M is mostly covered by the plot line of Online-P as both perform similarly.
The main difference between both approaches is handling uncorrelated relationships between input data size and task runtime.
Here, Online-M estimates a runtime according to the median, whereas Online-P considers a statistical distribution.
However, our data shows that nearly all tasks have correlated data, leading to similar results.

In Figure~\ref{fig:cumu-sub-first}, Lotaru has an $MPE \leq 10.00\%$ for 50\% of all combinations , while for Online-M and Online-P 50\% of all combinations have an $MPE \leq 21.60\%$.
The naive approach has the highest prediction error with more than 30.12\% of all combinations having an $MPE > 1$.

In comparison to this, Figure~\ref{fig:cumu-sub-second} exhibits slight changes between Online-M and Online-P.
Further, it shows a high slope at the beginning for Online-P, Online-M, and Lotaru.
In this region, all three approaches score similarly, however, higher prediction errors occur less frequently for Lotaru.

Regardless of our results, in the following experiments, we select all ten partitions for the prediction models to provide a fair comparison between Lotaru and the baselines.

\begin{figure*}
    \includegraphics[width=\textwidth]{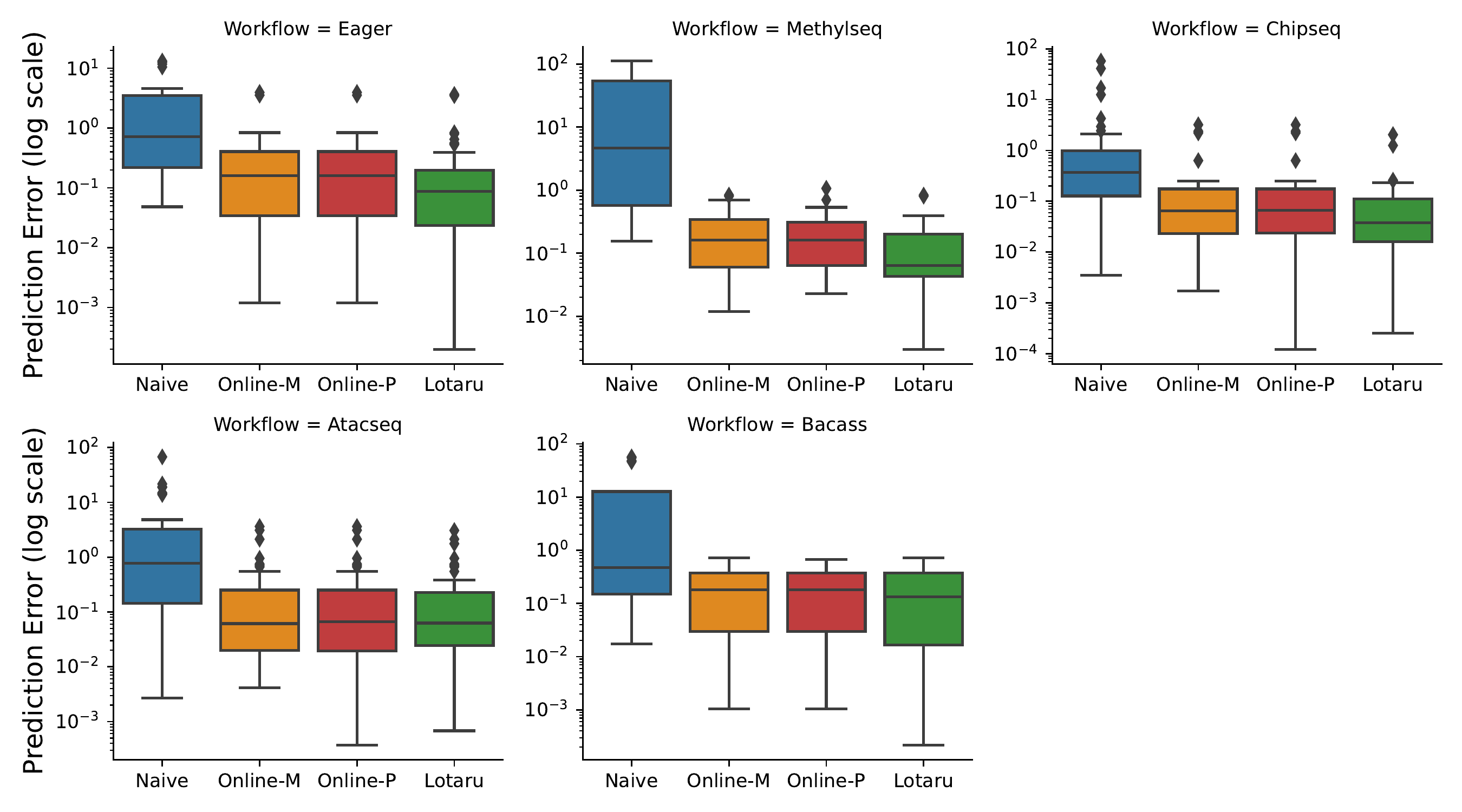}
    \caption{Prediction Errors for a homogeneous cluster where no model adjustments are required.} \label{fig:estimation_error_local}
\end{figure*}
\subsection{Predictions for a Homogeneous Cluster}

In our next experiment, we investigate the prediction performance without having to cope with heterogeneity in the cluster.
Therefore, we assume that the local machine is similar to the target nodes and that no model adjustment is needed.
Figure~\ref{fig:estimation_error_local} shows results for each workflow separately.
One can see that Lotaru outperforms the three baselines, achieving an MPE over all workflow tasks of 5.70\%, while the best performing baseline, Online-P, has an MPE of 10.34\%.
The largest difference can be observed for the Eager-2 workflow, where Lotaru achieves an MPE of 9.54\% while Online-P results in an MPE of 19.40\%.
An exception is the Ataqseq-1 workflow, where Online-M and Online-P can achieve a lower MPE of 4.27\% compared to Lotaru's 6.03\%, however, our max error is 55.00\% lower.
In three out of five workflows, Lotaru achieves a lower max error, while for two workflows similar max error values are achieved.

\begin{table}[]
\centering
\caption{Median difference between the actual factor and the factor calculated through Lotaru for Eager-1.}
\begin{tabular}{|c|c|c|c|c|c|}
\hline
Node                     & A1   & A2   & N1   & N2   & C2   \\ \hline
Median Factor Difference & 0.15 & 0.14 & 0.17 & 0.06 & 0.03 \\ \hline
\end{tabular}
\label{tab:facDiff}
\end{table}

\begin{table*}[]
\centering
\caption{Comparison of the calculated adjustment factor and the actual runtime factor between the local and the C2 machine for all tasks of the Eager-1 workflow.}
\begin{tabular}{|c|r|r|r|r|r|r|r|}
\cline{1-8}
Task       & bwa    & bcftools\_stats & samtools\_f\_a\_f & damageprofiler & preseq   & genotyping\_hc   & fastqc\_a\_c   \\ \cline{1-8}
Actual Factor     & 0.87   & 0.79            & 0.88              & 0.82           & 0.84     & 0.83             & 0.92           \\ \cline{1-8}
Calculated Factor & 0.87   & 0.87            & 0.87              & 0.87           & 0.87     & 0.87             & 0.87           \\ \cline{1-8}
\arrayrulecolor{white}\cmidrule[0.1pt]{1-8}\arrayrulecolor{black}
\cline{1-7}
Task       & fastqc & samtools\_flag  & samtools\_filter  & markduplicates & qualimap & adapter\_rem             \\ \cline{1-7}
Actual Factor    & 0.90   & 0.84            & 0.65              & 0.86           & 0.76     & 0.83                             \\ \cline{1-7}
Calculated Factor & 0.87   & 0.87            & 0.88              & 0.88           & 0.86     & 0.87                             \\ \cline{1-7}
\end{tabular}
\label{tab:adjustment}
\end{table*}%

\subsection{Model Adjustment} A critical point in Lotaru's approach is the mapping of the local predictions to the heterogeneous target nodes.
With the model adjustment, we want to adapt the predicted runtime for hardware differences on the target machines.
First, Table~\ref{tab:facDiff} compares the differences between the actual runtime factor and the factor Lotaru calculated.

The term difference refers to the absolute difference between actual factor and calculated factor.
One can see that for C2 and N2 the difference is the lowest.
This is expected since our infrastructure profiling showed that these nodes are the closest to the local node regarding performance characteristics.
In contrast, a difference of 0.14 or 0.15 for A1 and A2 seems high, however, both machines have a significant difference in the actual hardware characteristics and thus, only score half of the CPU events/s and much lower I/O values.

For Table~\ref{tab:adjustment}, we continue to take the Eager-1 workflow as an example.
The table compares the estimated adjustment factor with the actual factor between the local machine and the C2 machine for all 13 tasks in the Eager-1 workflow.
One can see that for only three out of 13 tasks, the difference is greater than 0.05, while the median difference between our estimated factor and the actual ratio is 0.03.

The task fastqc from Table~\ref{tab:adjustment} is a common bioinformatics task that spots potential problems in sequences data and occurs in all five workflows.
Therefore, we chose the task fastqc for our analysis across all machines.

The actually calculated factor for fastqc on C2 shows a difference of only 0.03 and N2 shows an even lower difference of 0.02.
Similarly as in Table~\ref{tab:facDiff}, again, the outlier is node A1.
Node A1 yields a difference of 0.31, whereby one has to consider that the actual factor of 2.37 is much higher than for the other machines.
Therefore, an absolute difference of 0.31 for an actual factor of 2.37 corresponds to a relative difference of 13.08\%. 

Concluding, our used adjustment strategy is able to accurately reflect hardware differences.

\subsection{Predictions for a Heterogeneous Cluster}
\begin{figure*}
    \includegraphics[width=\textwidth]{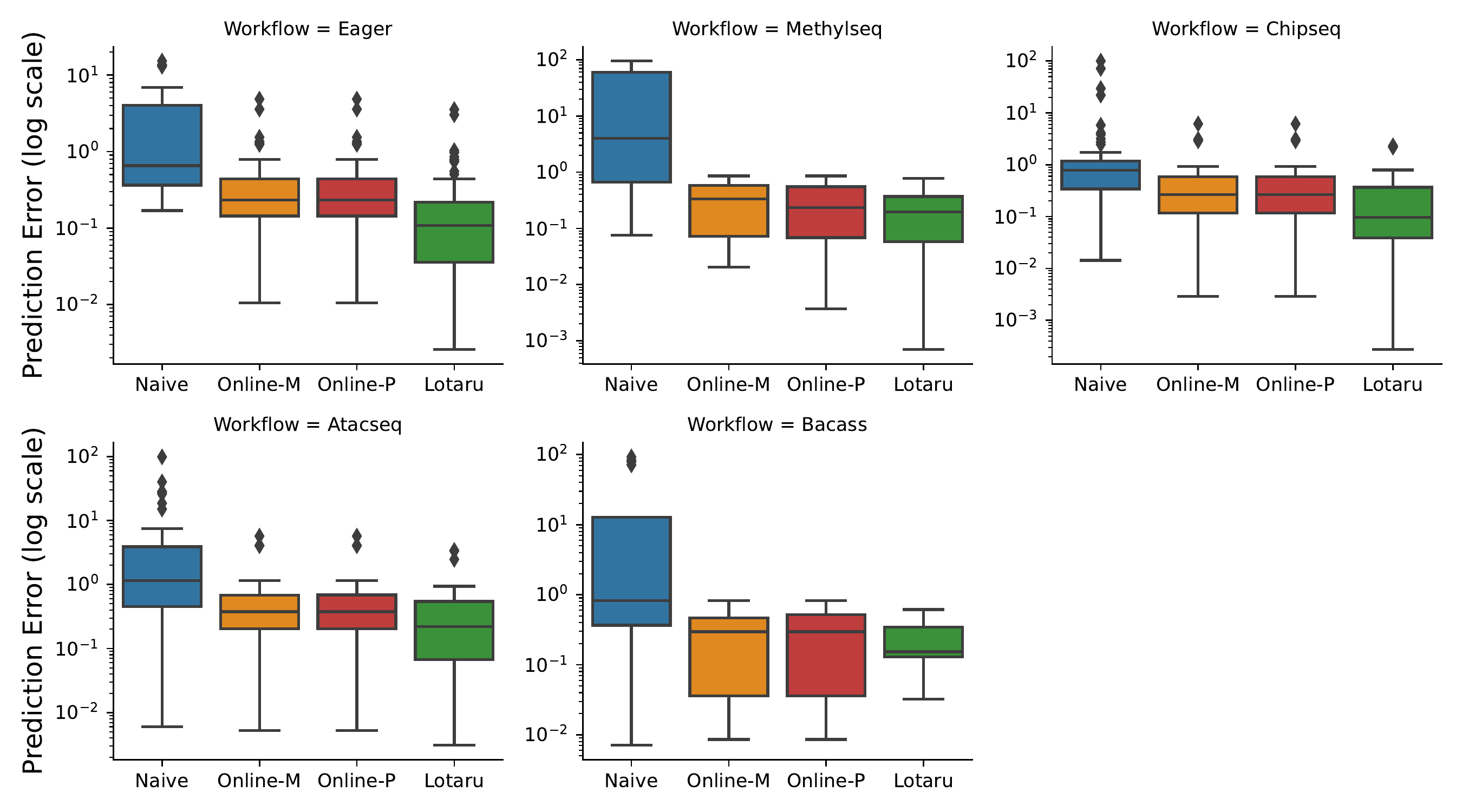}
    \caption{Prediction Errors for target node C2 with the influence of our model adjustment, logarithmic scale.} \label{fig:estimation_error_c2}
\end{figure*}

\begin{table}[]
\centering
\caption{MPE for all approaches on all machines over the five experiment workflows.}
\begin{tabular}{|c|c|c|c|c|c|c|}
\hline
Node     & A1    & A2    & N1    & N2    & C2     \\ \hline
Naive    & 53.11\% & 52.65\% & 58.53\% & 73.01\% & 83.10\%  \\ \hline
Online-M & 41.82\% & 39.96\% & 20.21\% & 18.40\% & 30.58\% \\ \hline
Online-P & 41.82\% & 39.91\% & 20.20\% & 18.40\% & 30.43\%  \\ \hline
Lotaru   & \textbf{21.71\%} & \textbf{19.91\%} & \textbf{14.19\%} & \textbf{13.80\%} & \textbf{14.62\%}   \\ \hline
\end{tabular}
\label{tab:predHet}
\end{table}

In the third experiment, we use our local machine and predict the runtimes for all target nodes A1, A2, N1, N2, and C2 for all tasks in all five evaluation workflows. 
Table~\ref{tab:predHet} gives an overview of the median prediction errors from all four prediction approaches.
Over all workflow tasks and the five target nodes, Lotaru achieves a median prediction error of 15.99\% compared to 30.90\% for Online-P, which constitutes a prediction error reduction of 48.25\%.
The differences in prediction errors are thus much more considerable for the more realistic case of heterogeneous clusters than for a homogeneous cluster.

Further, one can see that for N2, the node closest to the local machine regarding profiled performance characteristics, the estimation error of Lotaru and Online-P is the lowest for all target machines.
The error increases for both approaches once the machines differ more, i.e., the error for A1 and A2 is much higher than for N1 or N2.
However, while the differences regarding the prediction error between Lotaru and Online-P for N1 and N2 are rather small, they increase for A1 and A2.

Figure~\ref{fig:estimation_error_c2} gives more insights into the prediction errors for the example target node C2. 
Again, Lotaru outperfoms all three baselines. 
The best one, Online-P, achieves an MPE of 30.43\% compared to 14.62\% for Lotaru.
Lotaru's 75th percentile is always below the mean of the other predictors and two times below the median.
Regarding the 90th and 95th percentile, Lotaru consistently achieves the lowest prediction error.
Additionally, our predictions yield a lower standard deviation for all workflows.
Figure~\ref{fig:estimation_error_c2} shows that for four out of the five workflows, Lotaru's minimum prediction error is lower than the prediction error of any baseline.
In three out of five workflows the minimum prediction error is even three times lower. 
The maximum errors are always below the maximum values of any baseline.

\section{Conclusion}\label{sec:CONCLUSION}
This paper presented Lotaru, a system that estimates the runtime of scientific workflow tasks on the developer's machine.
To this end, Lotaru profiles the target infrastructure with microbenchmarks, reduces the input data to execute the workflow locally quickly, and estimates the runtime with a Bayesian regression based on the gathered data points.
We presented an implementation of our approach, while the code and the experimental data have been made available online.
Further, our introduced interface is easily extendable to other domains with other input data like images in remote sensing or astrophysics.

Lotaru works for all workflows that consist of multiple input files, which are, at least partly, processed separately and can be downsampled.
In contexts where such a downsampling is not possible, Lotaru could also use different subsets of the input files at the price of a somewhat longer training phase.
Further, since most big data analysis tasks, such as scans, show a linear input-runtime relation, Lotaru also assumes this linear relation.

Our evaluation with five real-world bioinformatics workflows on different data inputs shows that Lotaru's local estimation approach achieves low prediction errors and outperforms classical runtime prediction baselines.
For predictions where the target node is equal to the machine where the local profiling ran, Lotaru achieves an median prediction error over all workflows of 5.70\%, while the best performing baseline achieves 10.34\%.
Our comparison for the runtime predictions on five target nodes that differ from the local machine shows a median error over all workflows of 15.99\% compared to 30.90\% for the best competitor, a reduction of the prediction error of 48.25\%.

In the future, we plan to extend our implementation to different domains and adopt existing schedulers to consider our prediction's confidence and uncertainty values.

\begin{acks}
Funded by the Deutsche Forschungsgemeinschaft (DFG, German Research Foundation) as FONDA (Project 414984028, SFB 1404).
\end{acks}

\bibliographystyle{ACM-Reference-Format}
\bibliography{paper}

\end{document}